\documentclass[useAMS,usenatbib]{mn2e}
\usepackage{epsfig}
\newcommand{\me}{{\rm e^{-}}}
\newcommand{\mH}{{\rm H}}
\newcommand{\Hp}{{\rm H}^{+}}
\newcommand{\Hm}{{\rm H}^{-}}
\newcommand{\mHtp}{{\rm H_{2}^{+}}}
\newcommand{\mHt}{{\rm H_{2}}}

\newcommand{\hii}{\hbox{H\,{\sc ii}}\,}
\def\simless{\mathbin{\lower 3pt\hbox
   {$\rlap{\raise 5pt\hbox{$\char'074$}}\mathchar"7218$}}}   
\def\simgreat{\mathbin{\lower 3pt\hbox  
   {$\rlap{\raise 5pt\hbox{$\char'076$}}\mathchar"7218$}}} 

\title[Radiative feedback from ionized gas]{Radiative feedback from ionized gas}
\author[S. C. O. Glover]{S. C. O. Glover$^{1}$\thanks{E-mail: sglover@aip.de} \\
$^{1}$Astrophysikalisches Institut Potsdam, An der Sternwarte 16, D-14482 Potsdam, Germany}

\begin{document}

\maketitle

\begin{abstract}
$\mHt$ formation in metal-free gas occurs via the intermediate $\Hm$ or
$\mHtp$ ions. Destruction of these ions by photodissociation therefore serves to
suppress $\mHt$ formation. In this paper, I highlight the fact that several 
processes that occur in ionized primordial gas produce photons energetic
enough to photodissociate $\Hm$ or $\mHtp$ and outline how to compute
the photodissociation rates produced by a particular distribution of ionized 
gas.  I also show that there are 
circumstances of interest, such as during the growth of \hii regions around 
the first stars, in which this previously overlooked form of radiative 
feedback is of considerable importance.
\end{abstract}

\begin{keywords} 
atomic processes -- astrochemistry -- galaxies: formation -- cosmology: theory
\end{keywords}

\section{Introduction}
It has long been known that ro-vibrational line emission from molecular hydrogen,
$\mHt$, is the dominant cooling process in gas of primordial composition at 
temperatures $200 < T < 10^{4} \: {\rm K}$ \citep{sz67,pd68,mst69,h69}. 
Moreover, recent work has made it clear that $\mHt$ cooling also plays a 
key role in the evolution of low-density metal-poor gas \citep{jgkm07}. 
In primordial gas and in low density metal-poor gas, $\mHt$ formation on 
dust is ineffective, and $\mHt$ forms primarily through the gas-phase 
reactions \citep{mc61, pd68}
\begin{eqnarray}
\mH + \me & \rightarrow & \Hm + \gamma, \\
\Hm + \mH & \rightarrow & \mHt + \me, \label{R1}
\end{eqnarray}
with a smaller contribution coming from \citep{sz67}
\begin{eqnarray}
\mH + \Hp & \rightarrow & \mHtp + \gamma, \\
\mHtp + \mH & \rightarrow & \mHt + \Hp. \label{R2}
\end{eqnarray}
Although only a small fractional abundance of $\mHt$, of order $10^{-3}$,
is produced by these reactions, many authors have shown that this is more
than sufficient to allow the gas to cool within a Hubble time, and to allow 
star formation to occur \citep[see e.g.][]{teg97,bcl02,abn02,yahs03}.

Once stars have formed, however, their radiation can interfere with
the gas-phase production of $\mHt$. Most of the attention devoted to the
study of so-called ``radiative feedback'' has focused on the ultraviolet 
radiation from massive stars, which can suppress further star formation 
by photoionizing atomic and molecular hydrogen \citep{htl96,kysu04,
wan04,wn06,awb07} and by 
photodissociating $\mHt$ \citep{hrl97,on99,har00,gb01}, with the latter 
process able to operate at far greater distances from the star, owing to 
the low opacity of the gas at frequencies redwards of the Lyman limit.  
On the other hand, radiative feedback due to the photodissociation of 
$\Hm$ and $\mHtp$
\begin{eqnarray}
\Hm + \gamma & \rightarrow & \mH + \me, \label{hmpd} \\
\mHtp + \gamma & \rightarrow & \mH + \Hp, \label{h2ppd}
\end{eqnarray}
has attracted little direct study. One reason for this is the speed
of reactions~\ref{R1} and \ref{R2}: both have rate coefficients 
$k \sim 10^{-9} \: {\rm cm^{3}} \: {\rm s^{-1}}$, and so for reactions
\ref{hmpd} or \ref{h2ppd} to be effective at suppressing $\mHt$ formation, 
they must occur at a rapid rate. The other main reason for the comparative
neglect of these processes is the nature of the radiation sources typically
assumed in studies of radiative feedback. Work to date has focused
on feedback from massive stars or from AGN. These sources are bright
in the far-UV, and previous studies have found that feedback due to
$\mHt$ photodissociation by UV photons from these sources becomes 
effective long before feedback due to $\Hm$ or $\mHtp$ photodissociation 
by optical or near-IR photons from the same sources \citep[see e.g.][]{mba01}.

However, one issue that does not appear to have been previously 
considered in any great detail is that fact that once sources of ionization 
such as massive stars or AGN exist, the ionized gas that they produce
will act as a secondary source of radiation. Radiative feedback due to 
the photodissociation of $\mHt$ by Lyman-Werner band photons 
produced by recombining ${\rm He^{+}}$ was considered by 
\citet{yokh06} and found to be unimportant. However, radiative feedback 
due to the photodissociation of $\Hm$ or $\mHtp$ by recombination 
emission or free-free emission from the ionized gas does not appear to 
have been considered prior to now.

In this paper, I investigate this form of feedback and show that there 
are circumstances of cosmological importance in which it can be highly
effective. The calculations presented in this paper involve a number
of simplifications, so as to facilitate the exploration of a variety of 
different scenarios with minimal computational effort. Nevertheless,
they should produce results that are correct to within an order of 
magnitude, and so serve to highlight which scenarios require more
detailed study.

The layout of this paper is as follows. In \S\ref{emiss}, I outline the
method used to compute the photodissociation rates produced by
a given distribution of ionized gas. In \S\ref{model}, I use this 
method to investigate the effects of $\Hm$ and $\mHtp$
photodissociation in the neutral gas surrounding an expanding \hii 
region created by a population III star (\S\ref{expand}),  in a neutral 
cloud embedded in a large \hii region (\S\ref{embed}) and in the
centre of a recombining `fossil' \hii region (\S\ref{fossil}).  Finally, 
in \S\ref{conclude} I summarize what these results tell us about the 
scenarios in which $\Hm$ and $\mHtp$ photodissociation is important.

\section{Modelling emission from ionized gas}
\label{emiss}
The radiation flux at a location $\mathbf{x}$ produced by emission from an ionized 
volume V is given by:
\begin{equation}
 F(\nu) = \frac{1}{4\pi} \int_{V} e^{-\tau(\nu, \mathbf{x^\prime}, \mathbf{x})}
 \frac{n_{\rm e}(\mathbf{x^\prime}) 
 \sum_{i} \gamma_{\rm i}(\nu; {\mathbf{x^\prime}}) 
 n_{\rm i}(\mathbf{x^\prime})}{|\mathbf{x^\prime} - \mathbf{x}|^{2}}
 \: {\rm d}\mathbf{x^\prime}, \label{emiss1}
\end{equation}
where $\tau(\nu, \mathbf{x^\prime}, \mathbf{x})$ is the optical depth
at frequency $\nu$ between $\mathbf{x}$ and $\mathbf{x^\prime}$,
$n_{\rm e}$ is the number density of electrons, $n_{\rm i}$
is the number density of ions of species $i$, $\gamma_{\rm i}$ is 
the emission coefficient for species $i$ 
(with units ${\rm erg} \: {\rm cm^{3}} \: {\rm s}^{-1} \: {\rm Hz^{-1}}$),
and where we sum over all ionic species present in the gas. For 
clarity, in the simplified examples presented in this paper, I restrict
my attention to emission from recombining $\Hp$ ions. In normal
circumstances this will produce the bulk of the emission, with a small
additional contribution coming from helium, and with only negligible
amounts of radiation coming from other ions. This simplification allows
us to rewrite equation~\ref{emiss1} as:
\begin{equation}
 F(\nu) = \frac{1}{4\pi} \int_{V} e^{-\tau(\nu, \mathbf{x^\prime}, \mathbf{x})}
 \frac{n_{\rm e}(\mathbf{x^\prime}) 
 n_{\rm H^{+}}(\mathbf{x^\prime}) 
\gamma_{\rm H^{+}}(\nu; {\mathbf{x^\prime}})}{|\mathbf{x^\prime} - \mathbf{x}|^{2}}
 \: {\rm d}\mathbf{x^\prime}, \label{emiss2}
\end{equation}
where $n_{\rm H^{+}}$ is the number density of $\Hp$ ions and where 
$\gamma_{\rm H^{+}}$ is the emission coefficient for emission from 
$\Hp$.  We can further simplify this equation by making the 
assumption that the temperature of the ionized gas is uniform. In this
case $\gamma_{\Hp}(\nu)$ becomes independent of position and can be
moved outside the integral:
\begin{equation}
 F(\nu) = \frac{\gamma_{\rm H^{+}}(\nu)}{4\pi} 
 \int_{V}  e^{-\tau(\nu, \mathbf{x^\prime}, \mathbf{x})}
 \frac{n_{\rm e}(\mathbf{x^\prime}) 
 n_{\rm H^{+}}(\mathbf{x^\prime})}{|\mathbf{x^\prime} - \mathbf{x}|^{2}}
 \: {\rm d}\mathbf{x^\prime}. \label{emiss3}
\end{equation}
A final simplification comes if we assume that the gas is optically thin.
At the densities of interest in this work, this is a good approximation 
for most of the emission, as the continuum opacity of metal-free gas is 
very small \citep{lcs91}. The exceptions are the bound-free emission 
produced by recombination directly into the $n = 1$ ground state, and 
the Lyman series lines. As I discuss at greater length in the next section,
this emission can be taken into account by assuming that case B 
recombination applies.

The assumption of optically thin gas allows us to rewrite 
Equation~\ref{emiss3} as:
\begin{equation}
 F(\nu) = \gamma_{\rm H^{+}}(\nu) I_{V},
 \end{equation}
 where
 \begin{equation}
 I_{V} = \frac{1}{4\pi} 
 \int_{V}  \frac{n_{\rm e}(\mathbf{x^\prime}) 
 n_{\rm H^{+}}(\mathbf{x^\prime})}{|\mathbf{x^\prime} - \mathbf{x}|^{2}}
 \: {\rm d}\mathbf{x^\prime}. 
\end{equation}
With these simplifications we have reduced the problem of computing the
flux into two separate, simpler problems: that of computing 
$\gamma_{\Hp}(\nu)$ and that of evaluating the integral $I_{V}$. 
For a given distribution of ionized gas, numerical
evaluation of the integral is trivial, and so the only real remaining 
difficulty is the evaluation of $\gamma_{\Hp}(\nu)$. I discuss this in 
the next section.

\subsection{Computing the emission coefficient}
\subsubsection{Bound-bound emission}
\label{bbems}
In my treatment of the effects of recombination, I assume that
Case B applies. In the classical Case B, the gas is optically
thick in both the Lyman continuum and the Lyman series 
lines. Therefore, emission produced by recombination directly
to the $n=1$ ground state will be immediately re-absorbed, 
with the result that only recombinations to excited states with 
$n \ge 2$ actually result in a net decrease in the number of 
ionized hydrogen atoms. Every recombination to a state with
$n \ge 2$ will be followed by a radiative cascade as the 
atom attempts to reach the ground state and some fraction
of the photons produced during these cascades will be capable 
of photodissociating $\Hm$ or $\mHtp$.

The photodissociation threshold of $\Hm$ is 0.755~eV,
and so $\Hm$ can be photodissociated by any photons
in the Lyman series or Balmer series, and by most photons 
in the Paschen series (starting with Paschen-$\beta$).
$\mHtp$, with its larger photodissociation threshold 
of 2.65~eV, can be dissociated by any Lyman series 
photon, and by most Balmer series photons (starting
with H$\gamma$), but is not affected by Paschen 
series photons. 

For the emissivities of the lines in the Balmer and Paschen 
series, I use the values computed for Case B by \citet{sh95},
who tabulate frequency-averaged emissivities for transitions 
from all excited states up to $n = 50$. As the widths of these
emission lines are much smaller than the frequency range 
over which the photodissociation cross-sections of $\Hm$ or 
$\mHtp$ vary substantially, accurate modelling of the line
profiles is unnecessary and so for simplicity I model the
lines as delta functions.

In Case B, all Lyman series photons more energetic than
Lyman-$\alpha$ are eventually degraded to Lyman-$\alpha$
photons plus one or more lower energy photons.
While the results of \citet{sh95} include these lower energy
photons, they do not account for the eventual fate of the 
Lyman-$\alpha$ photons. In the classical Case B, these 
photons can never escape from the ionized region: the 
optical depth of the Lyman series lines is assumed to be 
infinite. In reality, the optical depth is finite and some 
photons can escape from the gas, although generally 
not before they have scattered multiple times. There is 
also a small chance that an excited hydrogen atom in
the $2p$ state will reach the ground state not by direct
radiative decay to the $1s$ state, but rather by undergoing 
a collisional transition to the $2s$ state, followed by 
two-photon decay to the ground state. Over time, this 
process results in the net loss of a large number of 
Lyman-$\alpha$ photons from the ionized gas. 

The relative importance of these two processes depends 
upon a number of factors, such as the density of the \hii 
region and the Lyman-$\alpha$ optical depth of the gas. 
For simplicity, I consider in this paper only the limiting 
case in which two-photon decay dominates. In this case,
the contribution to $\gamma_{\Hp}(\nu)$ arising from 
two-photon emission can be written as \citep{fk06}
\begin{equation}
\gamma_{\rm 2ph}(\nu) =  \frac{2h\nu}{\nu_{\rm Ly\alpha}}
P(\nu/\nu_{\rm Ly\alpha}) \alpha_{\rm B}, 
\end{equation}
where $\nu_{\rm Ly\alpha}$ is the frequency at line center
of the Lyman-$\alpha$ line, $\alpha_{\rm B}$ is the Case B 
recombination coefficient and $P(y){\rm d}y$ is the 
normalized probability per two-photon decay of getting
one photon in the interval ${\rm d}y = {\rm d}\nu/\nu_{\rm Ly\alpha}$.
For $\alpha_{\rm B}$, I use the values tabulated by 
\citet{sh95}, while for $P(y)$ I use a fitting function taken from 
\citet{fk06}:
\begin{equation}
P(y) = 1.307 - 2.627 z^{2} + 2.563 z^{4} - 51.69 z^{6},
\end{equation}
where $z = y -0.5$.

\subsubsection{Bound-free and free-free emission}
\begin{figure}
\centering
\epsfig{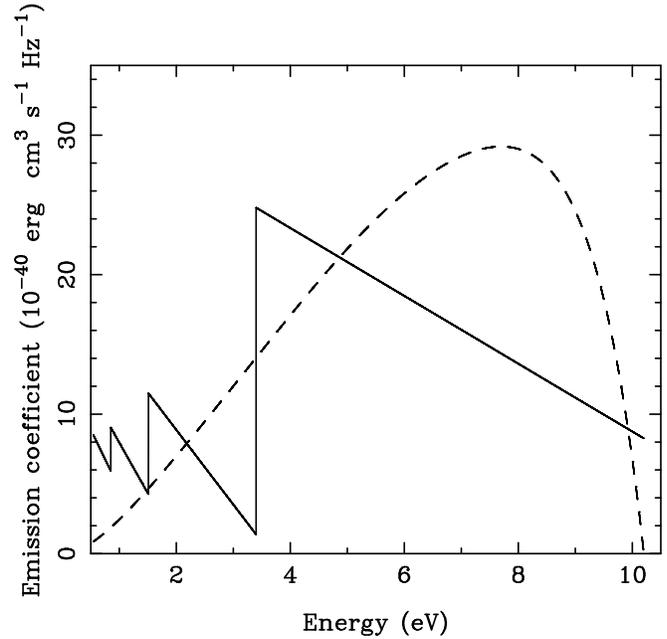}
\caption{Emission coefficients, plotted as a function of photon energy,
for the bound-free and free-free emission (summed and plotted as 
solid line) and two-photon emission (dashed line) produced by an
ionized gas of pure hydrogen with temperature $T = 10^{4} \: {\rm K}$.
\label{tot_emiss}}
\end{figure}

The contribution to $\gamma_{\Hp}(\nu)$  from bound-free and
free-free emission can be written as \citep{fk06}
\begin{equation}
\gamma_{\rm bf + ff} = \frac{6.84 \times 10^{-38}}{T^{1/2}} e^{-h\nu/kT}
\left[\bar{g_{\rm ff}} + \sum_{n=2}^{\infty} \frac{x_{\rm n}  e^{x_{n}}}{n} 
g_{\rm fb}(n)\right]
\label{bf+ff}
\end{equation}
where $T$ is the gas temperature, $x_{\rm n} =  1 {\rm Ryd} / (kTn^{2})$, 
$\bar{g_{\rm ff}}$ is the thermally averaged Gaunt factor for free-free 
emission and $g_{\rm fb}(n)$ is the Gaunt factor for free-bound 
emission from recombination into level $n$. The value of 
this expression has been tabulated as a function of frequency and
temperature by \citet{f80}. In Figure~\ref{tot_emiss}, I plot the value
of $\gamma_{\rm bf + ff}$ as a function of photon energy for photons
with $0.755 < h\nu < 13.6 \: {\rm eV}$, assuming a gas temperature 
$T = 10^{4} \: {\rm K}$. In the same figure, I also plot the two-photon
contribution $\gamma_{\rm 2ph}$. At low energies, close to the 
$\Hm$ photodissociation threshold, the bound-free and free-free
contributions dominate, while at high energies, the two-photon
contribution dominates. It is also plain from the figure that if we were
to disregard the two-photon emission, and to instead assume that
all of the Lyman-$\alpha$ photons produced by the ionized gas
eventually escape by scattering into the wings of the line, we would 
nevertheless obtain photodissociation rates for $\Hm$ and $\mHtp$
of the same order of magnitude as those derived below.

\subsection{Evaluating the photodissociation rates}
Given $\gamma_{\Hp}$, it is easy to compute the photodissociation
rates of $\Hm$ and $\mHtp$ as functions of the volume integral $I_{V}$.
The $\Hm$ photodissociation rate can be written as
\begin{eqnarray}
R_{\rm pd, \Hm} & = & \int_{0}^{\infty} \frac{\sigma_{\Hm}(\nu) F(\nu)}{h \nu} {\rm d}\nu, \\
 & = & I_{V}  \int_{0}^{\infty} \frac{\sigma_{\Hm}(\nu) \gamma_{\rm H^{+}}(\nu)}{h \nu} {\rm d}\nu
 \label{rate1}
\end{eqnarray}
where the photodissociation cross section is \citep{wis79}
\begin{equation}
\sigma_{\Hm}(\nu) = 7.928 \times 10^{5} \frac{(\nu - \nu_{\rm th})^{3/2}}{\nu^{3}}
\end{equation}
at frequencies above the $\Hm$ photodissociation threshold 
$h\nu_{\rm th} = 0.755 \: {\rm eV}$.  The $\mHtp$ photodissociation rate 
is given by a similar expression
\begin{eqnarray}
R_{\rm pd, \mHtp} & = & \int_{0}^{\infty} \frac{\sigma_{\mHtp}(\nu) F(\nu)}{h \nu} {\rm d}\nu, \\
 & = & I_{V}  \int_{0}^{\infty} \frac{\sigma_{\mHtp}(\nu) \gamma_{\rm H^{+}}(\nu)}{h \nu} {\rm d}\nu,
 \label{rate2}
\end{eqnarray}
where the photodissociation cross section at photon energies 
$11.27 \: {\rm eV} > h\nu > h\nu_{\rm th}$ is
\begin{equation}
\sigma_{\mHtp}(\nu) = {\rm dex} \left[-40.97 + 82.01 \eta -  93.22 \eta^{2} + 34.89 \eta^{3} \right],
\end{equation}
and at photon energies $h\nu > 11.27 {\rm eV}$ is
\begin{equation}
\sigma_{\mHtp}(\nu) = {\rm dex} \left[ -30.26 +  37.94 \eta - 34.03 \eta^{2} +  8.89 \eta^{3} \right],
\end{equation}
where $\eta$ is the photon energy in Rydbergs, and where the photodissociation
threshold energy $h\nu_{\rm th} = 2.65 \: {\rm eV} = 0.195 \: {\rm Ryd}$ \citep{d68}.

\begin{table}
\centering
\begin{tabular}{llccc}
\hline
 & &\multicolumn{3}{c}{\hii region temperature (K)} \\
 & & 5000 & 10000 & 20000 \\
 \hline
 & Continuum & 3.49 & 1.78 & 1.01 \\
$\Hm$ & Line & 23.8 & 3.44 & 0.720 \\
 & Total & 27.3 & 5.21 & 1.73 \\
\hline
 & Continuum  & 0.100 & 0.048 & 0.024 \\
$\mHtp$ & Line & $1.0 \times 10^{-10}$ & $3.0 \times 10^{-11}$ & $9.7 \times 10^{-12}$ \\
 & Total & 0.100 & 0.048 & 0.024 \\
\hline
\end{tabular}
\caption{Value of $R_{\rm pd, i} / I_{V}$, in units of 
$10^{-29} \: {\rm cm^{5}} \: {\rm s^{-1}}$, for $i = \Hm$ and 
$i = \mHtp$, computed for three different \hii region 
temperatures. The size of the contributions from continuum
emission and line emission is noted. \label{tab:rates}}
\end{table}

The values of the frequency integrals in equations~\ref{rate1} and \ref{rate2} are 
independent of the spatial distribution of the ionized gas, but do depend on its 
temperature. For gas temperatures in the range $3000 < T < 30000 \: {\rm K}$,
their values are fit to within 5\% by the expressions:
\begin{equation} 
\frac{R_{\rm pd, \Hm}}{I_{V}} = {\rm dex} \left[-28.28 - 2.04 \, t_{4} + 1.28 \, t_{4}^{2} \right]
\, {\rm cm^{5}} \, {\rm s^{-1}}
\end{equation}
and
\begin{equation}
\frac{R_{\rm pd, \mHtp}}{I_{V}}  =  {\rm dex} \left[-30.33 - 1.01 \, t_{4} + 0.24 \, 
t_{4}^{2} \right] \, {\rm cm^{5}} \, {\rm s^{-1}},
\end{equation}
where $t_{4} = \log (T / 10^{4} \: {\rm K})$. Table~\ref{tab:rates} shows 
the relative size of the line and continuum contributions for three different gas 
temperatures: $T = 5000 \: {\rm K}$, $10^{4} \: {\rm K}$ and 
$2 \times 10^{4} \: {\rm K}$. Two main points are worthy of note. Firstly, there 
is an obvious temperature dependence: the hotter the \hii region, the lower the 
photodissociative flux. Secondly, line emission is far more important in the case of $\Hm$ 
photodissociation that in the case of $\mHtp$ photodissociation. In
the former case, line emission is the dominant contribution 
when the temperature of the ionized gas is small and is significant 
even when $T$ is large. In the latter case, the contribution from line 
emission is always negligible. This difference is a consequence of the 
fact that the $\mHtp$ photodissociation cross-section is very small 
compared to the $\Hm$ photodissociation cross-section at the 
energies which correspond to Balmer series photons. The main
contribution to $\mHtp$ photodissociation comes from photons 
with energies $\sim 6 \: {\rm eV}$ or above and hence in this case 
the continuum processes dominate.

\section{Results}
\label{model}
In order to establish whether or not the photodissociaton of $\Hm$ and
$\mHtp$ by emission from ionized hydrogen is ever an important process,
I compute in this section the amount of flux produced in several scenarios
of cosmological interest, along with the size of the resulting photodissociation
rates.  As my aim in this paper is simply to produce estimates of these values
with the correct order of magnitude, the model \hii regions considered here 
are all highly simplified compared to realistic systems. Nevertheless, the 
results of  these simple models should be a good guide as to whether more
detailed numerical modelling is justified. 

\subsection{Expanding $\hii$ region}
\label{expand}
The first simple scenario considered here involves emission from an 
expanding \hii region within a small protogalactic halo. Because the
first stars to form in the Universe are predicted to be very massive
\citep{abn02,bcl02,yoha06,g06} and therefore to be emitters of 
large numbers of ionizing photons \citep{sch02}, this scenario 
has been studied numerically by a number of authors
\citep{wan04,wn06,abs06,jgb06,su06,awb07}. 
This work has given us a good understanding of the basic chain
of events. The nascent \hii region initially grows very rapidly, but
recombinations in the dense gas surrounding the ionizing source
quickly cause its growth to slow. After less than $10^{5} \: {\rm yr}$
(assuming a $100 \: M_{\odot}$ star), the ionization front bounding
the \hii region has become a slow, D-type front that drives a shock
ahead of itself into the dense neutral gas. At radii greater than the
shock radius $r_{\rm S}$, the density profile of the gas remains 
almost the same as before the switch-on of the ionizing source.
Within the \hii region, on the other hand, the density profile is flat,
indicating that the density of the ionized gas is almost constant.
At this point, the evolution of the \hii region is dominated by the
pressure-driven expansion of the ionized gas: as the gas expands,
the recombination rate falls, and so more ionizing photons become
available for ionizing previous neutral gas, expanding the amount
of mass contained in the \hii region. If the density profile of the 
protogalactic gas is steep enough -- it must fall off with radius more
quickly than $\rho \propto r^{-3/2}$ \citep{fttb90} -- and if the massive
star shines for long enough, then eventually the ionization front 
will undergo a transition back to a supersonic R-type front, and 
will soon thereafter break out of the protogalaxy. However, even
if the source switches off  before break-out occurs, most of the
ionized gas will nevertheless escape from the protogalaxy, as 
the mean outward radial velocity of this material is typically 
several times higher than the escape velocity of the halo.

Although the strong Lyman-Werner band emission from the massive 
star will quickly dissociate diffuse $\mHt$ throughout much of the 
halo \citep{on99,gb01}, the presence of abundant free electrons in the 
partially ionized region ahead of the ionization front can catalyze 
$\mHt$ formation in this region, and \citet{su06} have shown that 
the shielding provided by this $\mHt$ can allow $\mHt$ to survive
in dense cores elsewhere in the protogalaxy. However, these
calculations did not take into account the possible effects of 
recombination emission from the ionized gas.

\begin{figure}
\centering
\epsfig{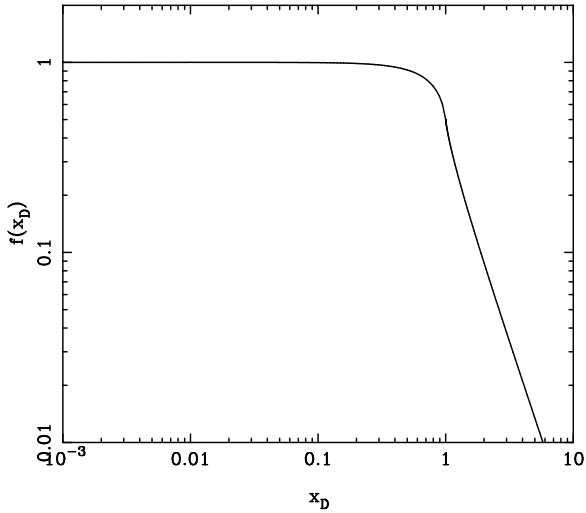}
\caption{Value of the function $f(x_{\rm D})$, defined in Equation~\ref{fx},
for a wide range of  $x_{\rm D}$. \label{intg_term}}
\end{figure}

To compute the emission from a typical protogalactic \hii region, we assume, 
for simplicity, that it is spherically symmetric, with radius $R_{\rm I}$, and that 
the gas within it has a uniform density $n_{\rm I}$. Inspection of the results from 
detailed three-dimensional numerical simulations of \hii region growth in 
these protogalactic systems suggests that these should be reasonable 
approximations. In that case, the value of $I_{V}$ at a distance $D > R_{\rm I}$
from the center of the \hii region is given by
\begin{equation}
I_{V} = n_{\rm I}^{2} R_{\rm I} f(x_{\rm D}),
\end{equation}
where
\begin{equation}
f(x_{\rm D}) = \int_{-1}^{1} \frac{1}{4} \ln \left[1 
+ \frac{1 - x^{2}}{(x_{\rm D} - x)^{2}} \right] \: {\rm d}x,  \label{fx}
\end{equation}
and where $x_{\rm D} = D / R_{\rm I}$. In deriving this expression I have assumed
that the gas is fully ionized, with $n_{\rm H^{+}} = n_{\rm e} = n_{\rm I}$. The value of 
$f(x_{\rm D})$ is plotted in Figure~\ref{intg_term} as a function of $x_{\rm D}$. In 
the limit that $x_{\rm D} \rightarrow 0$, we see that $I_{V} \rightarrow
n_{\rm I}^{2} R_{\rm I}$, while for $x_{\rm D} = 1$, $I_{V} = 
0.5 n_{\rm I}^{2} R_{\rm I}$ and in the limit of large $D$, 
$I_{V} \simeq (1 / 3D^{2}) n_{\rm I}^{2} R_{\rm I}$. 

If we assume that the temperature of the ionized gas is $2 \times 10^{4} \: {\rm K}$,
then at a point just outside the \hii region, the $\Hm$ photodissociation 
rate is given by
\begin{equation}
R_{\rm pd, \Hm} = 2.67 \times 10^{-11} \left(\frac{n_{\rm I}}{1 \: {\rm cm^{-3}}}\right)^{2}
\left(\frac{R_{\rm I}}{1 \: {\rm pc}}\right) \: {\rm s^{-1}},
\end{equation}
and the $\mHtp$ photodissociation rate is given by
\begin{equation}
R_{\rm pd, \mHtp} = 3.76 \times 10^{-13} \left(\frac{n_{\rm I}}{1 \: {\rm cm^{-3}}}\right)^{2}
\left(\frac{R_{\rm I}}{1 \: {\rm pc}}\right) \: {\rm s^{-1}}.
\end{equation}
For comparison, the rate at which $\Hm$ is destroyed by associative detachment
(reaction~\ref{R1}) is given by
\begin{equation}
R_{\rm c, \Hm} = 1.3 \times 10^{-9} n_{\mH} \: {\rm s^{-1}},
\end{equation}
where $n_{\mH}$ is the number density of atomic hydrogen, and where
we have adopted the rate coefficient determined for this reaction 
by \citet{sff67}. As noted by \citet{gsj06}, the rate of this reaction is uncertain
by almost an order of magnitude, but for the initial order-of-magnitude 
study given in this paper, the use of a value from close to the middle of 
the range of uncertainty is probably justified. The rate at which $\mHtp$ is 
destroyed by charge transfer (reaction~\ref{R2}) is given by \citep{KAR79}
\begin{equation}
R_{\rm c, \mHtp} = 6.4 \times 10^{-10} n_{\mH} \: {\rm s^{-1}}.
\end{equation}

In this particular example, the density of atomic gas just outside the \hii
region will be similar to the density of the ionized gas within the \hii
region, implying that $n_{\mH} \simeq n_{\rm I}$. Therefore, $\Hm$
will be destroyed primarily by photodissociation, rather than associative
detachment, if
\begin{equation}
n_{\rm I} \simgreat 50 \left(\frac{R_{\rm I}}{1 \: {\rm pc}}\right)^{-1} \: {\rm cm^{-3}}.
\end{equation}
Similarly, photodissociation will dominate the destruction of $\mHtp$ if
\begin{equation}
n_{\rm I} \simgreat 2400 \left(\frac{R_{\rm I}}{1 \: {\rm pc}}\right)^{-1} \: {\rm cm^{-3}}.
\end{equation}
How do these numbers compare with the results of the numerical
simulations of \hii region growth mentioned above? To take one
particular example, consider the \hii region simulated by \citet{awb07}.
At $t = 0.1 \: {\rm Myr}$ after the switch-on of the central ionizing
source, the \hii region in their simulation has a mean density 
$n_{\rm I} \sim 500 \: {\rm cm^{-3}}$ and a size $R_{\rm I} \sim 1 \: {\rm pc}$.
At a later time, $t = 1 \: {\rm Myr}$, its size has grown to 
$R_{\rm I} \sim 20 \: {\rm pc}$, while its mean density has fallen 
to $n_{\rm I} \sim 10 \: {\rm cm^{-3}}$. In both cases, the diffuse flux 
from the \hii region is strong enough to dominate the destruction
of $\Hm$ in the surrounding gas. It does not dominate the destruction
of $\mHtp$, but is important at the 10--20\% level. Only once 
$t = 2.7 \: {\rm Myr}$ (at which point $R_{\rm I} \sim 50 \: {\rm pc}$ and
$n_{\rm I} \sim 1 \: {\rm cm^{-3}}$) does the diffuse flux decrease to a 
level at which it no longer dominates the destruction of $\Hm$.
However, by this point in the evolution of the protogalaxy, the
expansion of the central \hii region has accelerated almost all
of the surrounding gas to velocities greater than the protogalactic
escape velocity. Therefore, any $\mHt$ that does manage to
form via $\Hm$ at $t > 2.7 \: {\rm Myr}$ will not be retained by 
the protogalaxy.

Although the preceding analysis is for gas located just outside
the \hii region, $\Hm$ photodissociation will actually dominate
out to much larger distances. This can be easily seen if we 
compare the dependence on radius of $R_{\rm pd, \Hm}$ and
$R_{\rm c, \Hm}$. In the limit of large distances,  $R_{\rm pd, \Hm}$ 
scales with the distance $D$ as $R_{\rm pd, \Hm} \propto D^{-2}$,
while at smaller distances the scaling is less steep. If the density 
distribution of the neutral gas is given by $n \propto D^{-\alpha}$,
then $R_{\rm c, \Hm} \propto D^{-\alpha}$. Therefore, if $\alpha > 2$, 
then $R_{\rm c, \Hm}$ will fall off more steeply with increasing radius 
than $R_{\rm pd, \Hm}$. As simulations of population III star 
formation generally find that $\alpha \simeq 2.2$ 
\citep[see e.g.][]{yoha06}, this implies that if $\Hm$ photodissociation
dominates near the \hii region, it will also dominate throughout
the remainder of the protogalaxy.

\subsection{Neutral cloud embedded in large \hii region}
\label{embed}
The second scenario considered in this paper is that of a cloud of neutral 
hydrogen embedded in a large cosmological \hii region. This arrangement
of ionized and neutral gas may be encountered within the earliest 
protogalaxies after the formation of the first stars if any clumps of gas 
exist within these protogalaxies
that are dense enough to survive the passage of the expanding 
ionization front (I-front) without being immediately photoevaporated 
\citep{su06,s07}. On larger scales, this arrangement can also be used to 
represent primordial minihalos that have been enveloped by a large
intergalactic \hii region, but that have not yet been completely 
photoionized or photoevaporated \citep[see e.g.][]{isr05,as07}.

For simplicity, I assume that both the neutral cloud and the surrounding \hii 
region are spherically symmetric, with radii $R_{\rm cl}$ and $R_{\rm I}$ 
respectively. The cloud is assumed to have a power-law density profile
\begin{equation}
n(R) = n_{\rm cl} \left(\frac{R_{\rm cl}}{R}\right)^{\alpha} \label{sph_cl}
\end{equation}
with $\alpha < 3$, where $n_{\rm cl}$ is the density of the cloud at 
$R = R_{\rm cl}$, which can be written in terms of the cloud mass 
$M_{\rm cl}$ as
\begin{equation}
n_{\rm cl} = \frac{M_{\rm cl}}{m_{\rm H}} \frac{3 - \alpha}{4 \pi R_{\rm cl}^{3}},
\end{equation}
where $m_{\rm H}$ is the mass of a hydrogen atom. At early times, 
the outer edge of the cloud will be photoionized, but will not yet 
have had time to dynamically respond to the increased pressure, 
and so will maintain the same density profile as the neutral gas. 
The density of ionized gas at radii $R_{\rm cl} > R > R_{\rm cl, I}$,
where $R_{\rm cl, I}$ is the innermost extent of the ionized region, 
is therefore given by Equation~\ref{sph_cl} above, while at radii
$R \geq R_{\rm cl}$ I take the ionized gas density, $n_{\rm I}$, to be 
constant. Finally, it is necessary to specify the distance $D$ of 
the centre of the cloud from the centre of the \hii region.  

These assumptions allow us to write the volume integral $I_{V}$ as
\begin{eqnarray}
I_{V} &  = &    \frac{n_{\rm cl}^{2} R_{\rm cl}}{2\alpha - 1} 
 \left[\left(\frac{R_{\rm cl}}{R_{\rm cl, I}}\right)^{2\alpha - 1} - 1\right] \nonumber \\
 & + &  n_{\rm I}^{2} R_{\rm I} f(x_{\rm D}) - n_{\rm I}^{2} R_{\rm cl},
 \end{eqnarray}
where $x_{\rm D} = D/R_{\rm I}$ and where $f(x_{\rm D})$ is defined
by Equation~\ref{fx}. The first term in this expression corresponds to the
flux produced by the ionized outer edge of the cloud, while the
second and third terms, taken together, correspond to the flux from the 
surrounding constant density \hii region. 

We can explore the behaviour of this expression by considering a
concrete example. Let us suppose that $D = 0.5 R_{\rm I}$, so
that $x_{\rm D} = 0.5$, and that $R_{\rm cl, I} = 0.5 R_{\rm cl}$ and
$\alpha = 2$. Then $I_{V}$ becomes
\begin{equation}
I_{V} = \frac{7}{3} n_{\rm cl}^{2} R_{\rm cl} + 0.91 n_{\rm I}^{2} R_{\rm I}
 - n_{\rm I}^{2} R_{\rm cl}.
\end{equation}
If $R_{\rm I} \gg R_{\rm cl}$, then the second term dominates,
unless the density contrast between the cloud and the surrounding
\hii region is large. The first term dominates only if $n_{\rm cl} > 
0.6 (R_{\rm I}/R_{\rm cl})^{1/2} n_{\rm I}$, while the third
term is always negligible in these conditions. The required density 
contrast can be significantly reduced if we steepen the density profile 
of the cloud by increasing $\alpha$, or increase the fraction of it which 
is already ionized by decreasing $R_{\rm cl, I}$. Altering $D$, on the 
other hand, has little effect on the required contrast.

To proceed with our example, let us suppose that the \hii region
has a size $R_{\rm I} = 3 \: {\rm kpc}$ and number density 
$n_{\rm I} = 10^{-3} \: {\rm cm^{-3}}$ \citep{jgb06}, that 
the cloud has a virial radius $R_{\rm cl} = 100 \: {\rm pc}$,
and that $n_{\rm cl} = 0.2 \: {\rm cm^{-3}}$. In that case, the 
flux from the ionized edge of the cloud dominates, 
$I_{V} \simeq 2.9 \times 10^{19} \: {\rm cm^{-5}}$, and the
resulting $\Hm$ and $\mHtp$ photodissociation rates are
$R_{\rm pd, \Hm}  =  5.0 \times 10^{-10} \: {\rm s^{-1}}$
and $R_{\rm pd, \mHtp}  = 6.9 \times 10^{-12} \: {\rm s^{-1}}$
respectively. In comparison, 
$R_{\rm c, \Hm} = 1.04 \times 10^{-9} \: {\rm s^{-1}}$ and
$R_{\rm c, \mHtp} = 5.12 \times 10^{-10}  \: {\rm s^{-1}}$
at $R_{\rm cl, I}$. Therefore, in this particular example the
flux is of limited importance.  If we
were to assume that  $R_{\rm cl, I} = 0.05 R_{\rm cl}$,
rather than $R_{\rm cl, I} = 0.5 R_{\rm cl}$, however,
then the photodissociation rates would increase by a factor of a
thousand, while the collisional rates would increase
by only a factor of a hundred. In that case, more $\Hm$
would be destroyed by photodissociation than by associative
detachment. On the other hand,  the photodissociation of 
$\mHtp$ would still be of limited importance. 

This simple example demonstrates that in this situation, 
radiative feedback due to the emission from the ionized 
gas is most effective during the final stages of the 
photoionization of the neutral cloud, when the highest
density material is being photoionized. As we expect
dynamical effects to have become important by this 
stage in the photoionization process \citep{as07}, it is
difficult to assess the ultimate importance of the feedback
without performing more detailed calculations.

\subsection{Recombining fossil \hii region}
\label{fossil}
The final scenario examined here is that of a so-called `fossil' 
\hii region \citep{oh03}, i.e.\ an \hii region in which the ionized source 
has switched off. For simplicity, I consider the evolution of the gas
at the center of a spherically symmetric \hii region with constant 
density $n_{\rm I}$ and radius $R_{\rm I}$. If the gas is initially fully 
ionized, then $I_{V}$ at the moment of switch-off is given by
\begin{equation}
I_{V} = n^{2}_{\rm I} R_{\rm I}.
\end{equation}
If we again assume that the temperature of the ionized gas is 
$2 \times 10^{4} \: {\rm K}$, then the flux produced by the ionized
gas gives rise to photodissociation rates for $\Hm$ and $\mHtp$ 
that are given by
\begin{eqnarray}
R_{\rm pd, \Hm} & = & 5.3 \times 10^{-8} \left(\frac{n_{\rm I}}{1 \: {\rm cm^{-3}}} \right)^{2}
\left(\frac{R_{\rm I}}{1 \: {\rm kpc}} \right) \: {\rm s^{-1}}, \\
R_{\rm pd, \mHtp} & = & 7.4 \times 10^{-10} \left(\frac{n_{\rm I}}{1 \: {\rm cm^{-3}}} \right)^{2}
\left(\frac{R_{\rm I}}{1 \: {\rm kpc}} \right) \: {\rm s^{-1}}. 
\end{eqnarray}
As time passes, however, the gas in the \hii region will recombine.
Since $I_{V} \propto x^{2}$, where $x$ is the fractional ionization
of the gas, this means that both $R_{\rm pd, \Hm}$ and 
$R_{\rm pd, \mHtp}$ will both decrease with time. If the temperature
of the ionized gas were to be kept fixed, then these rates would also
fall off as $x^{2}$. In fact, it is likely that some cooling of the gas will
occur and so the effects of the decrease in $x$ will be offset to 
some extent by an increase in the emission coefficient. Nevertheless,
it is clear from Table~\ref{tab:rates} that changes in the emission coefficient 
will alter the rates by at most an order of magnitude, while a large
decrease in $x$ may alter them by many orders of magnitude. 
Therefore, consideration of the simplified case in which $T$ is kept
fixed serves to illustrate the basic behaviour of  the gas without
requiring one to model its thermal evolution. 

For a intergalactic \hii region with the same size and
density as in the previous section (i.e.\ $R_{\rm I} = 3 \: {\rm kpc}$
and $n_{\rm I} = 10^{-3} \: {\rm cm^{-3}}$), we therefore have
\begin{eqnarray}
R_{\rm pd, \Hm} & = & 1.59 \times 10^{-13} x^{2} \: {\rm s^{-1}}, \\
R_{\rm pd, \mHtp} & = & 2.22 \times 10^{-15} x^{2} \: {\rm s^{-1}}.
\end{eqnarray}
The destruction rate of $\Hm$ by associative detachment in the
same conditions is given approximately by
\begin{equation}
R_{\rm c, \Hm} = 1.3 \times 10^{-12} (1 - x) \: {\rm s^{-1}}.
\end{equation}
Clearly, $R_{\rm pd, \Hm} > R_{\rm c, \Hm}$ only if $(1 - x) \ll 1$.
However, if the fractional ionization is as high as this, then $\Hm$
will also be destroyed rapidly by mutual neutralization with protons
\begin{equation}
\Hm + \Hp \rightarrow \mH + \mH.
\end{equation}
The rate coefficient of this reaction is uncertain, possibly by as much
as an order of magnitude \citep{gsj06}. If we take the smallest of the
values quoted in the literature, the rate of \citet{dl87}, then we find that 
the mutual neutralization rate in our example \hii  region is
\begin{equation}
R_{\rm mn, \Hm} = 7 \times 10^{-10} T^{-1/2} x \: {\rm s^{-1}}.
\end{equation}
This is comfortably larger than $R_{\rm pd, \Hm}$ for all temperatures 
in our range of interest. Therefore, in this particular scenario, $\Hm$
photodissociation is unimportant. Comparison of the photodissociation
rate of $\mHtp$ with the rate at which it is destroyed by charge transfer
with $\mH$ (reaction~\ref{R2}) or by dissociative recombination
\begin{equation}
 \mHtp + \me \rightarrow \mH + \mH,
\end{equation}
leads to a similar conclusion regarding $\mHtp$.

Can we avoid these conclusions by considering a larger \hii region?
If we suppose that we have somehow managed to create a fossil
\hii region with $R_{\rm I} = 3 \: {\rm Mpc}$, rather than $3 \: {\rm kpc}$,
then it is easy to see that  the resulting photodissociation rates  would 
be a thousand times larger than in the case considered above. Even so, 
associative detachment still dominates the destruction of $\Hm$ if 
$x < 0.1$. Similarly, photodissociation is unimportant in comparison
 to dissociative recombination or charge transfer for $x < 0.5$. Since
 $\mHt$ formation in cooling, recombining gas becomes significant
 only once $x < 0.1$ \citep[see e.g.][]{ohh02}, it is clear that even in
 this somewhat unrealistic case, photodissociation of $\Hm$ and
 $\mHtp$ is unimportant.

The other way in which to increase the importance of photodissociation
is by increasing the density of the ionized gas. If, instead of a large 
intergalactic \hii region, we consider a small interstellar \hii region
with $n_{\rm I} = 100 \: {\rm cm^{-3}}$ and $R_{\rm I} = 10 \: {\rm pc}$,
then we find that
\begin{eqnarray}
R_{\rm pd, \Hm} & = & 5.3 \times 10^{-6} x^{2} \: {\rm s^{-1}}, \\
R_{\rm pd, \mHtp} & = & 7.4 \times 10^{-8} x^{2} \: {\rm s^{-1}}.
\end{eqnarray}
In this case, $R_{\rm pd, \Hm} > R_{\rm c, \Hm}$ only if $x > 0.15$,
while $R_{\rm pd, \mHtp} > R_{\rm c, \mHtp}$ only if $x > 0.59$,
so again the effects of photodissociation are unimportant. 

\section{Conclusions}
\label{conclude}
The results from the simple models  considered in the previous section 
demonstrate that in some circumstances, bound-free, free-free and 
two-photon emission from ionized gas can combine to produce a
significant $\Hm$ photodissociation rate. The effect is very sensitive
to the density of the ionized gas, since both the recombination rate 
and the free-free emission rate scale as the square of the density. 
However, the required densities are not excessively high and there 
are a number of situations in which we may encounter them. 

The particular examples considered here in which $\Hm$ 
photodissociation proves to be important are the early growth of an
 \hii region around a population III star, prior to its break-out from the
confining protogalaxy (\S\ref{expand}) and the late stages of
the photoionization of a neutral cloud (or minihalo) by an external radiation
source (\S\ref{embed}).  On the other hand, photodissociation of $\Hm$
is not important within so-called `fossil' \hii regions, owing to the rapid
decrease of the photodissociating flux associated with the decrease
in the fractional ionization of the recombining gas.

The destruction of $\mHtp$ by emission from the ionized gas proves
to be a less important effect, due to the much smaller photodissociation
rate (compared to $\Hm$) that results from the same amount of 
emission. The main reason for this is that the $\mHtp$ photodissociaton
cross-section is small (or zero) at the photon energies where much 
of the energy from the ionized gas is radiated, becoming significant
only for photons with energies greater than about $6 \: {\rm eV}$.
Nevertheless, even if $\mHtp$ survives where $\Hm$ does not, the
net effect is still a significant reduction in the $\mHt$ formation rate,
as $\mHt$ formation via $\mHtp$ occurs much more slowly than 
$\mHt$ formation via $\Hm$ \citep{gl03}.

\section*{Acknowledgments}
The author would like to thank T.\ Greif and the anonymous referee
for their comments on an earlier draft of this paper.


\begin{thebibliography}{}

\bibitem[Abel, Bryan, \& Norman(2002)]{abn02}
Abel, T., Bryan, G.~L., \& Norman, M.~L. 2002, Science, 295, 93

\bibitem[Abel, Wise \& Bryan(2007)]{awb07}
Abel, T., Wise, J.~H., Bryan, G.~L. 2007, ApJL, in press; astro-ph/0606019.

\bibitem[Ahn \& Shapiro(2007)]{as07}
Ahn, K., \& Shapiro, P.~R. 2007, MNRAS, 375, 881

\bibitem[Alvarez, Bromm \& Shapiro(2006)]{abs06}
Alvarez, M.~A., Bromm, V., \& Shapiro, P.~R. 2006, ApJ, 639, 621

\bibitem[Bromm, Coppi, \& Larson(2002)]{bcl02}
Bromm, V., Coppi, P.~S., \& Larson, R.~B. 2002, ApJ, 564, 23

\bibitem[Dalgarno \& Lepp(1987)]{dl87}
Dalgarno, A., \& Lepp, S. 1987, in Astrochemistry, ed.\ M.~S.\ Vardya 
\& S.~P.\ Tarafdar (Dordrecht: Reidel), 109

\bibitem[Dunn(1968)]{d68}
Dunn, G.~H. 1968, Phys.\ Rev., 172, 1

\bibitem[Ferland(1980)]{f80}
Ferland, G.~J. 1980, PASP, 92, 596

\bibitem[Fernandez \& Komatsu(2006)]{fk06}
Fernandez, E.~R., \& Komatsu, E. 2006, ApJ, 646, 703

\bibitem[Franco, Tenorio-Tagle, \& Bodenheimer(1990)]{fttb90}
Franco, J., Tenorio-Tagle, G., \& Bodenheimer, P. 1990, ApJ, 349, 126

\bibitem[Gao et~al.(2006)]{g06}
Gao, L., Abel, T., Frenk, C.~S., Jenkins, A., Springel, V., 
\& Yoshida, N. 2006, astro-ph/0610174

\bibitem[Glover \& Brand(2001)]{gb01}
Glover, S.~C.~O., \& Brand, P.~W.~J.~L. 2001, MNRAS, 321, 385

\bibitem[Glover(2003)]{gl03}
Glover, S.~C.~O. 2003, ApJ, 584, 331

\bibitem[Glover, Savin \& Jappsen(2006)]{gsj06}
Glover, S.~C.~O., Savin, D.~W., \& Jappsen, A.-K. 2006, ApJ, 640, 553

\bibitem[Haiman, Abel \& Rees(2000)]{har00}
Haiman, Z., Abel, T., \& Rees, M.~J. 2000, ApJ, 534, 11

\bibitem[Haiman, Rees \& Loeb(1997)]{hrl97}
Haiman, Z., Rees, M.~J.,\&  Loeb, A. 1997, ApJ, 476, 458
(Errata in ApJ, 484, 985)

\bibitem[Haiman, Thoul \& Loeb(1996)]{htl96}
Haiman, Z., Thoul, A., \& Loeb, A. 1996, ApJ, 464, 523.

\bibitem[Hirasawa (1969)]{h69}
Hirasawa, T., 1969, Prog.\ Theor.\ Phys., 42, 523

\bibitem[Hummer \& Storey(1987)]{hs87}
Hummer, D.~G., \& Storey, P.~J. 1987, MNRAS, 224, 801

\bibitem[Iliev, Shapiro \& Raga(2005)]{isr05}
Iliev, I., Shapiro, P.~R., \& Raga, A.~C. 2005, MNRAS, 361, 405

\bibitem[Jappsen et~al.(2007)]{jgkm07}
Jappsen, A.-K., Glover, S.~C.~O., Klessen, R.~S., \& {Mac Low}, M.-M.
2007, ApJ, in press; astro-ph/0511400

\bibitem[Johnson, Greif \& Bromm(2006)]{jgb06}
Johnson, J.~L., Greif, T.~H., \& Bromm, V. 2006, astro-ph/0612254

\bibitem[Karpas, Anicich \& Huntress(1979)]{KAR79}
Karpas, Z., Anicich, V., \& Huntress, W.~T. 1979, J.\ Chem.\ Phys, 70, 2877

\bibitem[Kitayama et~al.(2004)]{kysu04}
Kitayama, T., Yoshida, N., Susa, H., \& Umemura, M. 2004, ApJ, 613, 631

\bibitem[Lenzuni, Chernoff \& Salpeter(1991)]{lcs91}
Lenzuni, P., Chernoff, D.~F., \& Salpeter, E.~E. 1991, ApJS, 76, 759

\bibitem[Machacek, Bryan \& Abel(2001)]{mba01}
Machacek, M.~E., Bryan, G.~L., \& Abel, T. 2001, ApJ, 548, 509

\bibitem[Matsuda, Sato, \& Takeda(1969)]{mst69}
Matsuda, T., Sato, H., \& Takeda, H. 1969, Prog.\ Theor.\ Phys., 42, 219

\bibitem[McDowell(1961)]{mc61}
McDowell, M.~R.~C. 1961, Observatory, 81, 240

\bibitem[Oh \& Haiman(2002)]{ohh02}
Oh, S.~P., \& Haiman, Z. 2002, ApJ, 569, 558

\bibitem[Oh \& Haiman(2003)]{oh03}
Oh, S.~P., \& Haiman, Z. 2003, MNRAS, 346, 456

\bibitem[Omukai \& Nishi(1999)]{on99}
Omukai, K., \& Nishi, R. ApJ, 518, 64.

\bibitem[Peebles \& Dicke(1968)]{pd68}
Peebles, P.~J.~E., \& Dicke, R.~H. 1968, ApJ, 154, 891

\bibitem[Saslaw \& Zipoy(1967)]{sz67}
Saslaw, W.~C., \& Zipoy, D. 1967, Science, 216, 976

\bibitem[Schaerer(2002)]{sch02}
Schaerer, D. 2002, A\&A, 382, 28

\bibitem[Schmeltekopf et~al.(1967)]{sff67}
Schmeltekopf, A.~L., Fehsenfeld, F.~C., \& Ferguson, E.~E. 1967, ApJ, 118, L155

\bibitem[Storey \& Hummer(1988)]{sh88}
Storey, P.~J., Hummer, D.~G. 1988, MNRAS, 231, 1139

\bibitem[Storey \& Hummer(1995)]{sh95}
Storey, P.~J., Hummer, D.~G. 1995, MNRAS, 272, 41

\bibitem[Susa \& Umemura(2006)]{su06}
Susa, H., \& Umemura, M. 2006, ApJ, 645, L93

\bibitem[Susa(2007)]{s07}
Susa, H. 2007, astro-ph/0701172

\bibitem[Tegmark et~al.(1997)]{teg97}
Tegmark, M., Silk, J., Rees, M., Blanchard, A., Abel, T., \& Palla, F. 1997,
ApJ, 474, 1

\bibitem[Whalen, Abel \& Norman(2004)]{wan04}
Whalen, D., Abel, T., \& Norman, M.~L. 2004, ApJ, 610, 14

\bibitem[Whalen \& Norman(2006)]{wn06}
Whalen, D., Norman, M.~L., 2006, ApJS, 162, 281

\bibitem[Wishart(1979)]{wis79}
Wishart, A.~W. 1979, MNRAS, 187, 59P

\bibitem[Yoshida et~al.(2003)]{yahs03}
Yoshida, N., Abel, T., Hernquist, L., \& Sugiyama, N. 2003, ApJ,
592, 645

\bibitem[Yoshida et al.(2006a)]{yoha06}
Yoshida, N., Omuaki, K., Hernquist, L. \& Abel, T. 2006a, ApJ, 
652, 6

\bibitem[Yoshida et al.(2006b)]{yokh06}
Yoshida, N., Oh, S.~P., Kitayama, T., \& Hernquist, L. 2006b, ApJ, 
in press; astro-ph/0610819

\end{thebibliography}
\end{document}